\pdfoutput=1
\documentclass[aps,prd,showpacs,preprintnumbers,twocolumn]{revtex4-1}

\usepackage[colorlinks=true, pdfstartview=FitV, linkcolor=blue, citecolor=blue, urlcolor=blue]{hyperref}
\usepackage[dvips]{graphicx}
\usepackage{amsmath}
\usepackage{color}

\newcommand{\be}{\begin{equation}}
\newcommand{\ee}{\end{equation}}
\newcommand{\ba}{\begin{eqnarray}}
\newcommand{\ea}{\end{eqnarray}}

\begin{document}

\title{Quantum Monte Carlo simulation with a black hole}

\author{Sanjin Beni\'{c}} 
\affiliation{Physics Department, University of Zagreb, Zagreb 10000, Croatia}
\affiliation{Department of Physics, The University of Tokyo, Tokyo 113-0033, Japan}

\author{Arata~Yamamoto}
\affiliation{Department of Physics, The University of Tokyo, Tokyo 113-0033, Japan}

\date{\today}
\begin{abstract}
We perform quantum Monte Carlo simulations in the background of a classical black hole.
The lattice discretized path integral is numerically calculated in the Schwarzschild metric and in its approximated metric.
We study spontaneous symmetry breaking of a real scalar field theory.
We observe inhomogeneous symmetry breaking induced by inhomogeneous gravitational field.
\end{abstract}

\maketitle

\section{Introduction}

Quantum Monte Carlo method is the reliable computational scheme broadly used from condensed matter physics to elementary particles.
Although the method is conventionally formulated in flat spacetimes, it is also applicable to curved spacetimes \cite{Jersak:1996mn}.
We can study quantum phenomena in gravitational backgrounds by the quantum Monte Carlo method.

Black holes are an intriguing environment to explore the phenomena of symmetry breaking.
On the quantum level, the Hawking temperature of a black hole with radius $R=2GM$ \cite{Hawking:1974sw}
\begin{eqnarray}
\label{eqT}
  T = \frac{1}{4\pi R}
,
\end{eqnarray}
can trigger a phase transition provided it is higher from the critical transition temperature of some field theory.
Even when the Hawking temperature is lower than the critical temperature it has been suggested that symmetry may be restored near the horizon \cite{Hawking:1980ng}.
This can be important for the description of primordial black holes in the early universe, or micro black holes that could be created at particle colliders.
Symmetry breaking is also interesting in the context of vacuum polarization around compact stars \cite{Liddle:1989kh} and no-hair theorems \cite{Adler:1978dp}.
A related question concerns symmetry breaking in accelerated frames \cite{Hill:1985wi}.

In this work, we perform quantum Monte Carlo simulation of lattice scalar field theory in the presence of a black hole in thermal equilibrium.
The Compton wavelength of the particle is taken to be much larger than the black hole Schwarzschild radius.
We consider a real scalar field theory with spontaneously broken $Z_2$ symmetry.
We analyze inhomogeneous symmetry breaking induced by inhomogeneity of the spacetime.
It is known that the local temperature increases near black holes because of the Tolman-Ehrenfest effect \cite{Tolman:1930zza}. This will simply suppress symmetry breaking \cite{Hawking:1980ng}.
In quantum field theory, however, this is only one of the many possible effects.
A complete result is given by the competition among many effects.
Our finding is that symmetry breaking is strengthened close to the horizon.

The paper is organized as follows.
In Sec.~\ref{sec2}, we explain theoretical preliminaries in continuum theory.
In Sec.~\ref{sec3}, we present the formulation and results of lattice simulations.
Finally, Sec.~\ref{sec4} is devoted to the summary.

\section{Preliminaries}
\label{sec2}

Let us consider the real scalar field theory
\begin{equation}
\begin{split}
\label{eqS1}
S =& \ \int d^4x \sqrt{\det g(x)} \bigg[ \frac{1}{2} g^{\mu\nu}(x) \partial_\mu \phi(x) \partial_\nu \phi(x) \\
&+ \frac{1}{2} (m^2-\xi \mathcal{R}(x))\phi^2(x) + \frac{1}{4} \lambda \phi^4(x) \bigg]
,
\end{split}
\end{equation}
in a general coordinate $ds^2 = g_{\mu\nu}dx^\mu dx^\nu$ with the Euclidean signature $\det g > 0$.
Although the scalar curvature $\mathcal{R}$ affects symmetry breaking, black hole spacetime has $\xi \mathcal{R}=0$.
The $Z_2$ symmetry is spontaneously broken at low temperatures by tachyonic mass $m^2<0$.

We define a two-point function
\begin{equation}
\label{eqG}
 G(x,x') = \langle \phi(x) \phi(x') \rangle
.
\end{equation}
When the separation $x-x'$ is taken in the direction of the Killing vector, $G(x,x')$ is a function of $|x-x'|$, which is denoted by $G(|x-x'|)$.
(Although it depends on the coordinates in other directions, its dependence is omitted for simplicity.)
In the large separation limit, the two-point function gives the square of the condensate
\begin{equation}
 G(\infty) = \langle \phi \rangle^2
,
\end{equation}
(the off-diagonal long-range order) \cite{Yang:1962}.
In this work, we consider the condensate fraction
\begin{eqnarray}
\label{eqC1}
C = \frac{G(\infty)}{G(0)} = \frac{\langle \phi \rangle^2}{\langle \phi^2 \rangle},
\end{eqnarray}
as a dimensionless order parameter.
The physical interpretation of the condensate fraction becomes especially transparent in flat space. There it quantifies the ratio of number of condensed particles versus the total number of particles.
We will use the condensate fraction to quantify the strength of symmetry breaking in curved spacetimes.

Note that, in quantum field theory in curved spacetimes, the change of scale comes
from classical gravity and also from ultraviolet cutoff (e.~g., the lattice spacing in lattice regularization).
The scale is nontrivially modified by inhomogeneous renormalization in curved spacetimes.
Although the condensate fraction is dimensionless, it cannot eliminate this quantum correction.

\section{Lattice simulation}
\label{sec3}

We performed the conventional Monte Carlo simulation of real scalar field theory \cite{Montvay:1994cy}.
The scalar field action is regularized on the hypercubic lattice, and then the path integral with the lattice action is numerically calculated by the Monte Carlo sampling.

\begin{figure}[h]
 \includegraphics[width=.4\textwidth]{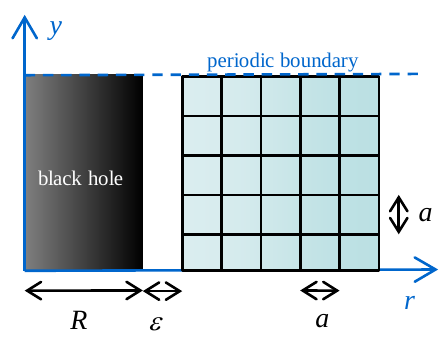}
 \caption{\label{fig1} 
 Geometry of the simplified coordinate \eqref{eqds1}.
 }
\end{figure}

Before considering the Schwarzschild coordinate, we consider the simplified coordinate
\begin{equation}
\begin{split}
\label{eqds1}
& ds^2 =  f(r) d\tau^2 +  \frac{1}{f(r)} dr^2 + dy^2 + dz^2
,\\
& f(r) = 1 - \frac{R}{r}
.
\end{split}
\end{equation}
This coordinate is derived by approximating the Euclidean Schwarzschild coordinate into the region $r \gg y,z$, namely, by neglecting the curvature in the $y$ and $z$ directions.
The lattice action is given by
\begin{equation}
\begin{split}
S =& \ \sum_x a^4 \bigg[ \frac{1}{2f(r)a^2} \{ \phi(x)-\phi(x-\hat{\tau}) \}^2
\\
&+ \frac{f(r)}{2a^2} \{ \phi(x)-\phi(x-\hat{r}) \}^2 
\\
&+ \frac{1}{2a^2} \{ \phi(x)-\phi(x-\hat{y}) \}^2
\\
&+ \frac{1}{2a^2} \{ \phi(x)-\phi(x-\hat{z}) \}^2
\\
&+ \frac{1}{2} m^2 \phi^2(x) + \frac{1}{4} \lambda \phi^4(x) \bigg]
,
\end{split}
\end{equation}
where $\hat\mu$ is the unit vector in $\mu$ direction.
The geometry is schematically shown in Fig.~\ref{fig1}.
There is a $(2+1)$-dimensional flat event horizon at $r=R$.
From Eq.~\eqref{eqT}, $R$ is given by
\begin{equation}
 R = \frac{1}{4\pi T} = \frac{N_\tau a}{4\pi}
.
\end{equation}
To avoid the coordinate singularity at $r=R$, we introduce the $r$ coordinates of lattice sites as $r = [R+\varepsilon, R+\varepsilon+(N_r-1) a]$, where $0<\varepsilon \ll a$.
We take free boundary conditions in the $r$ direction and periodic boundary conditions in the $y$, $z$, and $\tau$ directions.
We set $(ma)^2=-0.2$, $\lambda=0.2$, and $V = N_rN_yN_zN_\tau = 10 \times 10 \times 10 \times 60$.

\begin{figure}[h]
 \includegraphics[width=.5\textwidth]{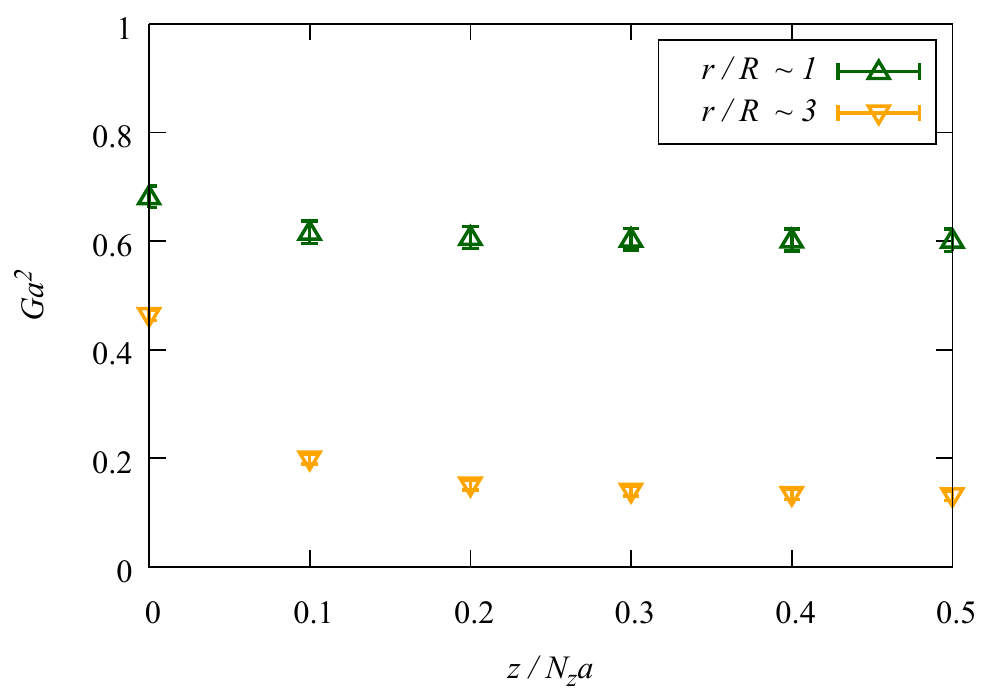}
 \caption{\label{figG1} 
 Two-point function $G(z)$ in the simplified coordinate \eqref{eqds1}.
 The data with $\varepsilon=0.1 a$ are shown.
 }
 \includegraphics[width=.5\textwidth]{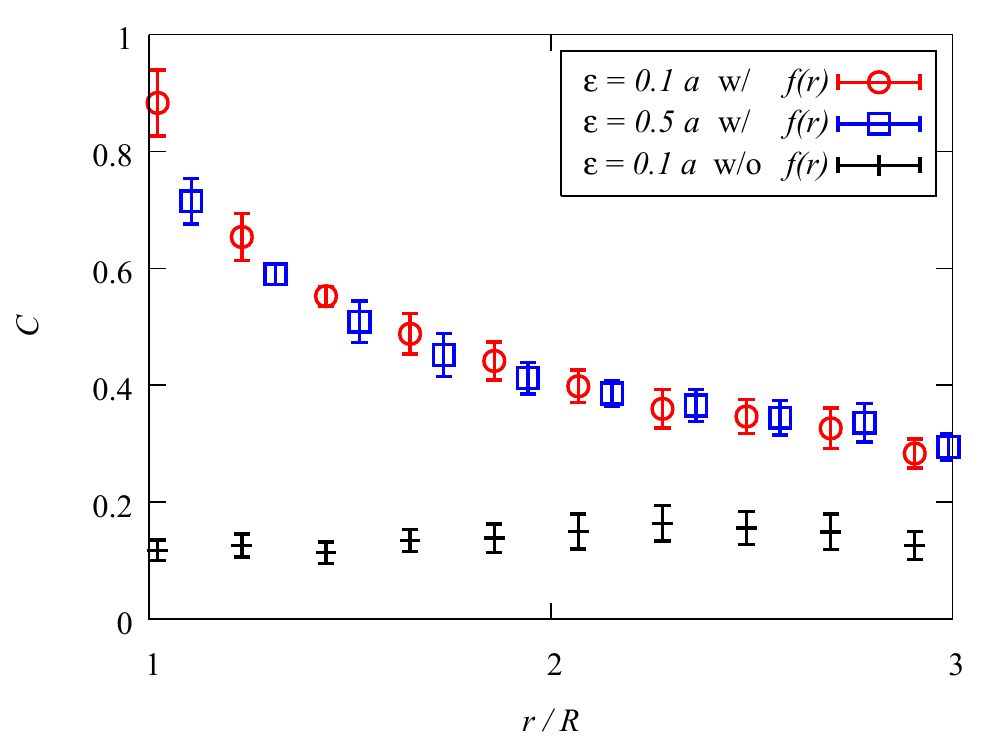}
 \caption{\label{figC1} 
 Condensate fraction $C(r)$ in the simplified coordinate \eqref{eqds1}.
 }
\end{figure}

On this lattice, we numerically calculate the two-point function in the $z$ direction.
As shown in Fig.~\ref{figG1}, we see clear plateaus indicating the off-diagonal long-range order.
We define the condensate fraction as
\begin{eqnarray}
C = \frac{G(N_za/2)}{G(0)}
,
\end{eqnarray}
because $N_za/2$ is the largest distance.
We show the $r$-dependence of the condensate fraction in Fig.~\ref{figC1}.
The calculation is performed for $\varepsilon=0.1 a$ and $\varepsilon=0.5 a$.
The results in both cases agree well with each other.
We find that the condensate fraction is enhanced by approaching the horizon $r/R = 1$.
We can attribute this enhancement to the gravitational redshift.
At the horizon $f(r)$ goes to zero and the coefficient of $(\partial_\tau \phi)^2$ in the action diverges.
As the coefficient of the derivative term becomes larger, a configuration with nonzero derivative has a large action and is thus disfavored.
This leads to a disfavoring of non-condensed configurations.
Consequently, the condensate fraction is enhanced.
In Fig.~\ref{figC1} we also show the results in the coordinate \eqref{eqds1} without $f(r)$, i.~e., in a flat spacetime.
The condensate fraction is finite and trivially independent of $r$ in the flat spacetime.

We numerically checked that symmetry is always preserved when $m^2=0$.
This is consistent with the above explanation.
The coefficient of the derivative term changes the magnitude of the condensate but does not trigger the tachyonic mass.

\begin{figure}[h]
 \includegraphics[width=.4\textwidth]{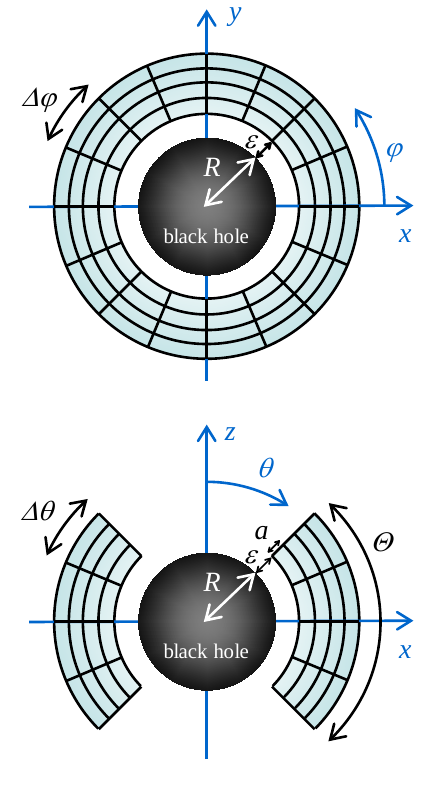}
 \caption{\label{fig2} 
 Geometry of the Schwarzschild coordinate \eqref{eqds2}.
 }
\end{figure}

Next, we consider the Euclidean Schwarzschild coordinate
\begin{equation}
\begin{split}
\label{eqds2}
& ds^2 =  f(r) d\tau^2 +  \frac{1}{f(r)} dr^2 + r^2 d\theta^2 + r^2 \sin^2 \theta d\varphi^2
,\\
& f(r) = 1 - \frac{R}{r}
.
\end{split}
\end{equation}
The lattice action is given by
\begin{equation}
\label{eqS2}
\begin{split}
S =& \ \sum_x a^2 \Delta \theta \Delta \varphi r^2 \sin \theta \bigg[ \frac{1}{2f(r)a^2} \{ \phi(x)-\phi(x-\hat{\tau}) \}^2
\\
&+ \frac{f(r)}{2a^2} \{ \phi(x)-\phi(x-\hat{r}) \}^2 
\\
&+ \frac{1}{2 r^2 \Delta \theta^2} \{ \phi(x)-\phi(x-\hat{\theta}) \}^2
\\
&+ \frac{1}{2 r^2 \sin^2 \theta \Delta \varphi^2} \{ \phi(x)-\phi(x-\hat{\varphi}) \}^2
\\
&+ \frac{1}{2} m^2 \phi^2(x) + \frac{1}{4} \lambda \phi^4(x) \bigg]
.
\end{split}
\end{equation}
with $\Delta \theta = \Theta / (N_\theta -1)$ and $\Delta \varphi = 2\pi / N_\varphi$.
We take the geometry shown in Fig.~\ref{fig2}.
There is a $(2+1)$-dimensional spherical event horizon at $r=R$.
To avoid the coordinate singularity at $r=R$ and $\sin \theta = 0$, we take $r = [R+\varepsilon, R+\varepsilon+(N_r-1) a]$ and $\theta = [\pi/2-\Theta/2, \pi/2+\Theta/2]$.
We take free boundary conditions in the $r$ and $\theta$ directions and periodic boundary conditions in the $\varphi$ and $\tau$ directions.
We set $(ma)^2=-0.2$, $\lambda=1$, $\Theta = \pi/2$, and $V = N_r N_\theta N_\varphi N_\tau = 10 \times 5 \times 16 \times 60$.

\begin{figure}[h]
 \includegraphics[width=.5\textwidth]{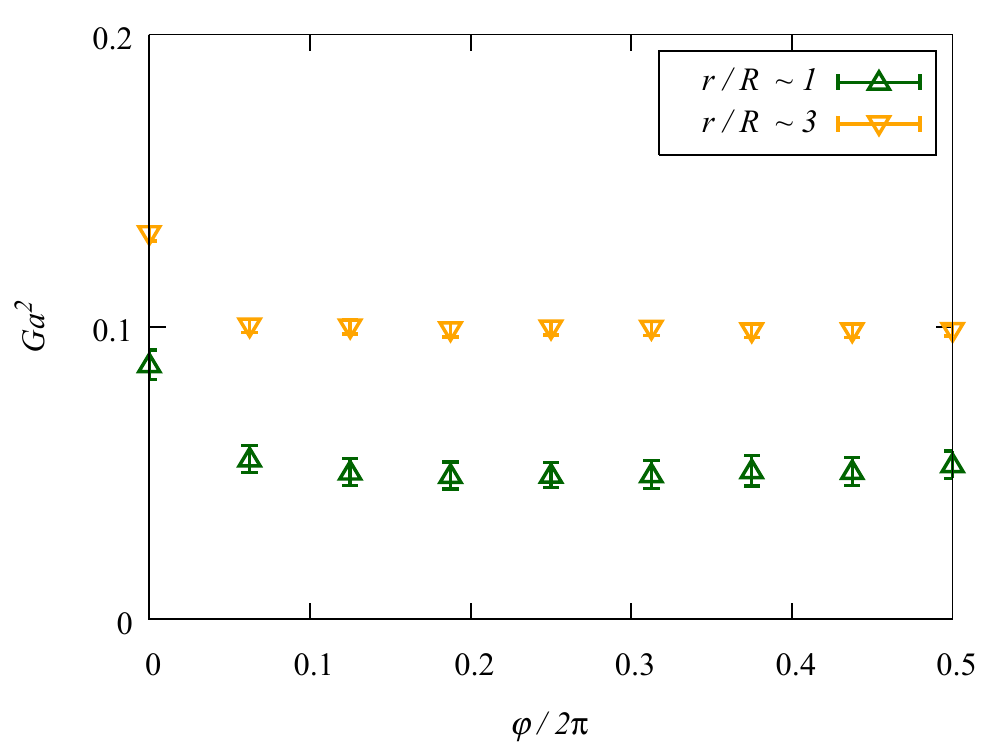}
 \caption{\label{figG2} 
 Two-point function $G(\varphi)$ in the Schwarzschild coordinate \eqref{eqds2}.
 The data with $\varepsilon=0.1 a$ and at $\theta=\pi/2$ are shown.
 }
 \includegraphics[width=.5\textwidth]{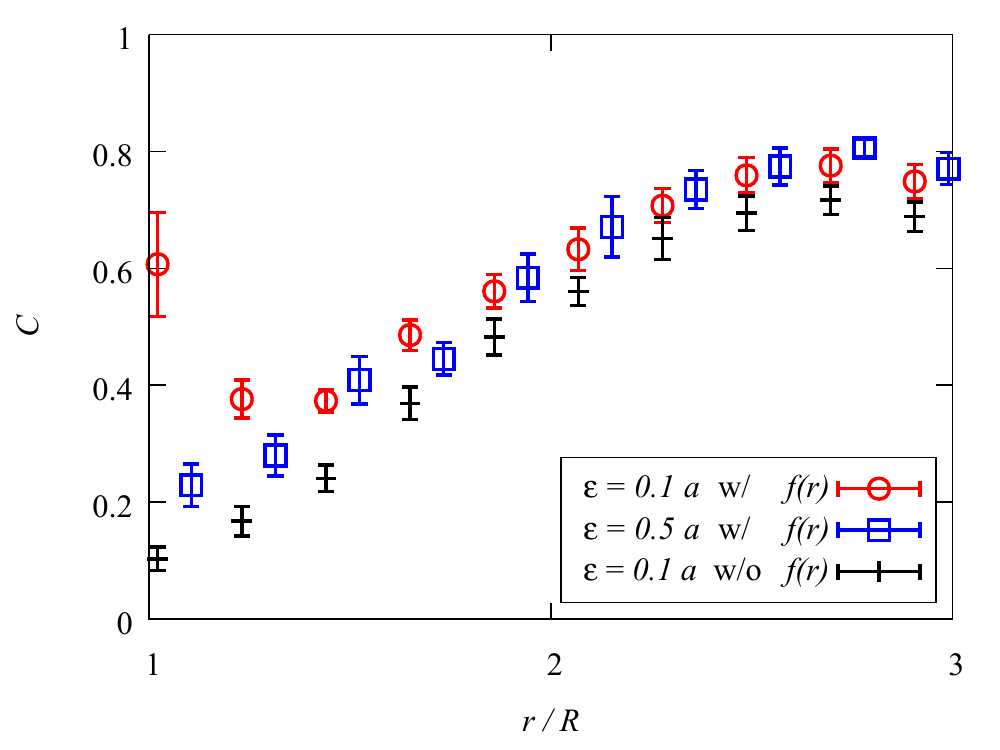}
 \caption{\label{figC2} 
 Condensate fraction $C(r)$ in the Schwarzschild coordinate \eqref{eqds2}.
 The data at $\theta=\pi/2$ are shown.
 }
\end{figure}

We calculated the two-point function in the $\varphi$ direction.
The condensate fraction is defined as
\begin{eqnarray}
\label{eqC2}
C = \frac{G(\pi)}{G(0)}
.
\end{eqnarray}
Although physical distance is finite in the  $\varphi$ direction, condensate is well-defined if non-condensate components are sufficiently small.
The results are shown in Figs.~\ref{figG2} and \ref{figC2}.
The result in the coordinate \eqref{eqds2} without $f(r)$, i.~e., in an ordinary spherical coordinate, is shown as a comparison.
The difference between the condensate fraction with $f(r)$ and without $f(r)$ increases near the event horizon.
This is essentially the same as in the simplified case.
Unlike the simplified case, the condensate fraction strongly depends on $r$ even without $f(r)$.
Since the transformation to a spherical coordinate must not change physics, this $r$-dependence is an artifact.
This is due to the strong $r$-dependence of the overall prefactor $\sqrt{\det g} = r^2 \sin \theta$.
This prefactor changes the unit cell $a^4$ in the Cartesian coordinate to $a^2 \Delta \theta \Delta \varphi r^2 \sin \theta$ in a spherical coordinate.
This change leads to the artificial $r$-dependence of physical parameters via renormalization.
Moreover, as seen in Fig.~\ref{figC2}, $\varepsilon$-independence is lost near the event horizon.

\begin{figure}[h]
 \includegraphics[width=.5\textwidth]{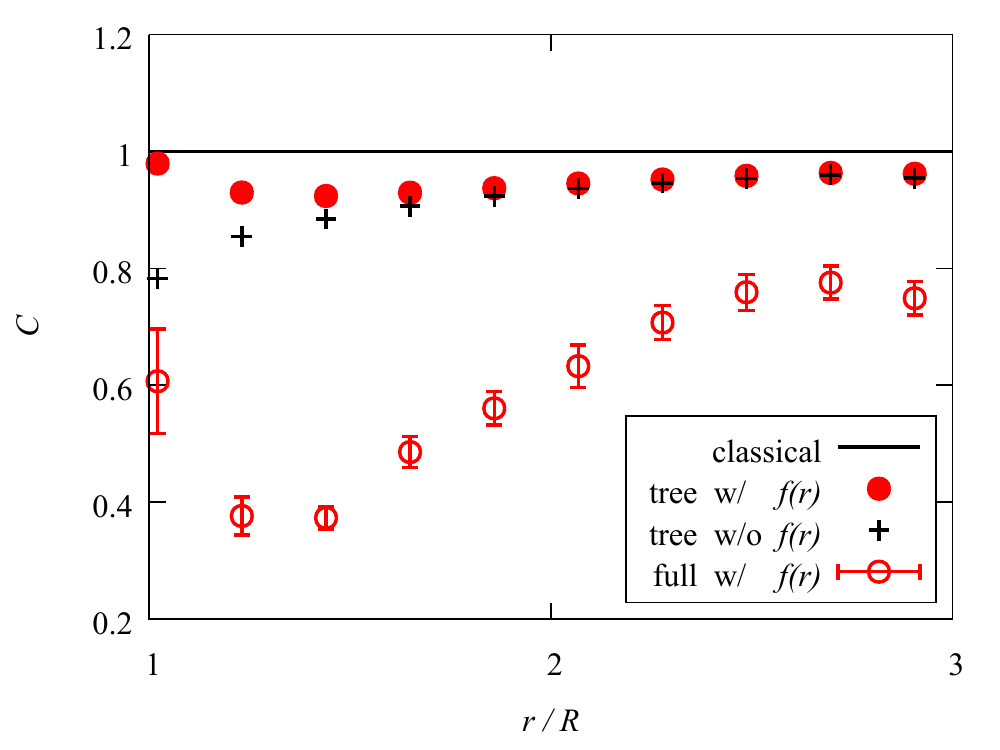}
 \caption{\label{figC3} 
 Classical solution, the tree-level results, and the full simulation result of the condensate fraction $C(r)$.
 The data with $\varepsilon=0.1 a$ and at $\theta=\pi/2$ are shown.
 }
\end{figure}

Finally, we compare the simulation with other calculations in Fig.~\ref{figC3}.
The classical condensate fraction is unity because the classical solution $\phi_c = \sqrt{-m^2/\lambda}$ is homogeneous and thus the two-point function is constant.
The tree-level result is the sum of the classical solution and the tree-level fluctuation around it.
For the detail, see Appendix.
The tree-level calculation is done with the same lattice action, parameters, and boundary conditions as the full simulation.
While the tree-level calculation and the full result are quantitatively different, we see the same qualitative behavior.
Even at the tree level, the condensate fraction without $f(r)$ depends on $r$.
This is because the tree-level fluctuation already possesses ultraviolet divergence and the result depends on the regularization to remove it.

\section{Summary}
\label{sec4}

We performed the first quantum Monte Carlo simulation with a black hole.
We considered a real scalar field theory with spontaneously broken $Z_2$ symmetry.
We found that spontaneous symmetry breaking is strengthened by redshift near black holes.
Unfortunately, the calculation in the Schwarzschild coordinates (essentially in the spherical coordinates) suffers from the artificial $r$-dependence of the regularization.
It is intuitively suggestive that the physical effect would be the difference of the condensate fraction with and without the redshift factor $f(r)$.
This is clearly supported by the calculation in the simplified coordinate.
However, for the practical use of the Schwarzschild coordinate, we need to solve the problem of the artificial $r$-dependence.
Actually, this is not a specific problem of lattice theory but a general problem of quantum field theory in curved spacetimes.
The theoretical calculation of quantum field theory has finite ambiguity stemming from regularization.
The ambiguity can be artificially inhomogeneous in curved spacetimes.
Physical inhomogeneity is hidden by such ambiguity.
Although it can be partially corrected by perturbative renormalization, the complete correction by nonperturbative renormalization is extremely difficult.

The application to chiral symmetry breaking in QCD is an interesting future work.
Although chiral symmetry is spontaneously broken in our present universe, it can be changed locally by black holes.
Not only redshift but also other various gravitational effects have been predicted for fermions \cite{Inagaki:1997kz}.
We can study the saga of chiral symmetry around black holes by evaluating the competition among them correctly in lattice QCD.

\begin{acknowledgments}
The authors thank Kenji Fukushima for useful discussions.
S.~B. was supported by the European Union Seventh Framework Programme (FP7 2007-2013) under grant agreement No. 291823, Marie Curie FP7-PEOPLE-2011-COFUND NEWFELPRO Grant No. 48.
A.~Y.~was supported by JSPS KAKENHI Grant Number 15K17624. 
The numerical simulations were carried out on SX-ACE in Osaka University.
\end{acknowledgments}

\appendix*

\section{Perturbative calculation}

Here, we provide the calculation of the lattice perturbation theory \cite{Montvay:1994cy}.
Since the Fourier transformation to momentum is ineffective in curved spacetimes, we work in coordinate space.
The lattice spacing is omitted in the following equations.

In a symmetric vacuum, the loop expansion is expressed by the free massless lattice propagator $D^{-1}_{x,y}$ and the second derivative of the potential
\be
M_{x,y} = \sqrt{\det g(x)} \{ m^2+ 3\lambda \phi^2(x) \} \delta_{x,y}
.
\ee
Now we consider the loop expansion in a broken vacuum.
The scalar field is shifted as
\begin{equation}
 \phi(x) = \phi_c + \Phi(x),
\end{equation}
where $\phi_c = \sqrt{-m^2/\lambda}$ is a classical solution of the action.
We rewrite
\be
D + M = A + B~,
\ee
where
\ba
A_{x,y} &=& D_{x,y} -2m^2 \sqrt{\det g(x)} \delta_{x,y}
\\
B_{x,y} &=& 3\lambda \sqrt{\det g(x)}  \{ 2\phi_c \Phi(x) + \Phi^2(x) \} \delta_{x,y}
.
\ea
These are $V \times V$ square matrices.
The expansion by the massless singular matrix $D$ becomes the expansion by the massive regular matrix $A$.

The two-point function of the original field $\phi$ is given by
\be
\begin{split}
\label{eqGloop}
G(x,y)
& = \langle \phi(x) \phi(y) \rangle
\\
& = \phi_c^2 + \phi_c[\Phi(x)+\Phi(y)] + \left[ \frac{\delta^2 \Gamma}{\delta \Phi(x)\delta \Phi(y)} \right]^{-1}
,
\end{split}
\ee
where $\Gamma$ is the perturbative effective action
\be
\begin{split}
\Gamma
&= S + \frac{1}{2}{\rm tr}{\rm Log}(A+B)~
\\
&= S + \frac{1}{2} {\rm tr}{\rm Log}A + \frac{1}{2} \sum_{n} \frac{1}{n} (-1)^{n-1} {\rm tr}[ (A^{-1}B)^n ]
.
\end{split}
\ee
The second term in Eq.~\eqref{eqGloop} does not contribute to the tree level.
At the tree level, we get the two-point function
\be
G(x,y)=  \phi_c^2 + A^{-1}_{x,y},
\ee
and the condensate fraction
\be
C = \frac{G(\pi)}{G(0)} = \frac{ \phi_c^2 + A^{-1}_{x,x+\pi\hat{\varphi}} }{\phi_c^2 + A^{-1}_{x,x}}
.
\ee

\end{document}